\documentstyle[aps,prl,floats]{revtex}
\begin{document}
\draft
\input epsf.sty

\twocolumn[\hsize\textwidth\columnwidth\hsize\csname 
@twocolumnfalse\endcsname
\title{Optically Pumped NMR
 Measurements of the Electron
 Spin Polarization
in GaAs Quantum Wells near
Landau Level Filling Factor {\boldmath $\nu$} = $\frac{\bf 1}{\bf 3}$}
\author{P. Khandelwal$^{1}$, N.\,N. Kuzma$^{1}$, S.\,E. Barrett$^{1}$, 
L.\,N. Pfeiffer$^{2}$, and K.\,W. West$^{2}$}
\address{$^{1}$Department of Physics, Yale University, New Haven,
 Connecticut 06511 \\ $^{2}$Bell Laboratories, Lucent Technologies, Murray
 Hill, New Jersey 07974}
\date{\today}
\maketitle
\begin{abstract}
 The Knight shift 
of $^{71}$Ga nuclei is measured in
two different electron-doped multiple quantum well samples
using optically pumped NMR.
 These data are the first direct 
 measurements of the electron
 spin polarization, ${\cal P}(\nu,T) \equiv
\frac{\langle S_{z}(\nu,T)\rangle}{\text{max}\langle S_{z}\rangle}$,
near $\nu$=$\frac{1}{3}$.  The ${\cal P}(T)$ data 
at $\nu$=$\frac{1}{3}$ 
  probe the neutral spin-flip excitations of a fractional
quantum Hall ferromagnet.  In addition,
 the saturated ${\cal P}(\nu)$
drops on either side of $\nu$=$\frac{1}{3}$, even 
 in a $B_{\text{tot}}$=12~Tesla field. 
The observed 
depolarization is quite small, consistent with an average
of $\sim 0.1$~spin-flips per quasihole (or quasiparticle), a
value which does not appear to be explicable by the current
theoretical understanding of the FQHE near $\nu$=$\frac{1}{3}$.
\end{abstract}

\vskip 2pc ] 

\narrowtext

The electron spin played no role in the earliest theory\cite{laughlin}
of the fractional quantum Hall effect (FQHE)\cite{fracnu},
 where the Zeeman 
energy $E_{Z}\,$$\equiv\,$$g^{\ast}\mu_{e}
B_{\text{tot}}$ was assumed to be infinite.
  However, for a two-dimensional electron system
(2DES) in GaAs, $E_{Z}$ is only $\sim$$\frac{1}{70}$ of the
electron-electron Coulomb energy
$E_{C}\,$$\equiv\,$$e^{2}/ \epsilon\, l_{B}\,$$\sim\,$$160\,$K
at 10~Tesla, raising the possibility
that interactions can lead to quantum Hall states with non-trivial spin
configurations\cite{halperin}.  
This idea underlies the recent theoretical predictions\cite{sondhi,fertig}
that the charged excitations of the $\nu$=1 integer quantum Hall ground state 
are novel spin-textures called skyrmions,
with experimentally observable consequences\cite{opnmr,science,skyrme,skyrme2}
 (Here $\nu$$\equiv$$n/n_B$, where $n$
is the number of electrons per unit area, and  
 $n_{B}\,$$=\,$$eB/hc\:$$\equiv\,$$1/2\pi l_{B}^{2}$ 
is the number of states per unit area in each Landau level).  
The spin physics near fractional $\nu$ should be even
more interesting, since it is the interactions that 
give rise to the FQHE\cite{tapash,prange,perspectives}.

In this Letter, we report optically pumped nuclear magnetic 
resonance (OPNMR)\cite{opnmrchar} studies of the 
Knight Shift $K_S$ 
of $^{71}$Ga  nuclei in two different electron-doped multiple
quantum well (MQW) samples. The $K_S$ data are the first 
direct observations of the spin polarization ${\cal P}(\nu,T)
\equiv
\frac{\langle S_{z}(\nu,T)\rangle}{\text{max}\langle S_{z}\rangle}$
of a 2DES near $\nu$=$\frac{1}{3}$.  These thermodynamic measurements
provide new insights into the physics of this important FQHE ground
state.

Both of the MQW samples in this study  were grown
 by molecular beam epitaxy on semi-insulating GaAs(001) substrates.
  Sample  40W contains forty $300\,$\AA\space wide GaAs wells separated
 by $3600\,$\AA\space wide Al$_{0.1}$Ga$_{0.9}$As barriers.   Sample 10W 
contains ten
$260\,$\AA\space wide wells separated by  $3120\,$\AA\space wide 
 barriers.  Silicon delta-doping spikes located in 
the center of each barrier provide the electrons that are confined in each GaAs
well at low temperatures, producing 2DES with very high mobility
($\mu > 1.4\times10^{6}$~cm$^{2}$/Vs).  This MQW structure also results in
a 2D electron density that is unusually insensitive to light, and extremely
uniform from well to well\cite{insensitive}. 
 The low temperature ($0.29\,$K $<$$\,T$$\,<\,$$20\,$K)
OPNMR measurements described below were performed using either a 
sorption-pumped~$^{3}$He cryostat or a $^{4}$He bucket dewar, in fields 
up to 12~Tesla.  The samples, about \mbox{
$4\,$$\times\,$6$\,$mm$^{2}$} in size, 
were in direct contact with helium, mounted
on the platform of a rotator assembly in the NMR probe.  Data were acquired
using the previously described\cite{opnmr,science} OPNMR timing sequence:
\mbox{SAT--$\tau_{L}$--$\tau_{D}$--DET}, modified for use 
below 1 Kelvin  (e.g., $\tau_{D}\sim\,$40~s,
laser power $\sim\,$10~mW/cm$^{2}$, low rf voltage levels). A calibrated
RuO$_{2}$ thermometer, in good thermal contact with the sample, was used to
record the temperature during signal acquisition.

\begin{figure}
\centerline{\epsfxsize=2.95in\epsfbox{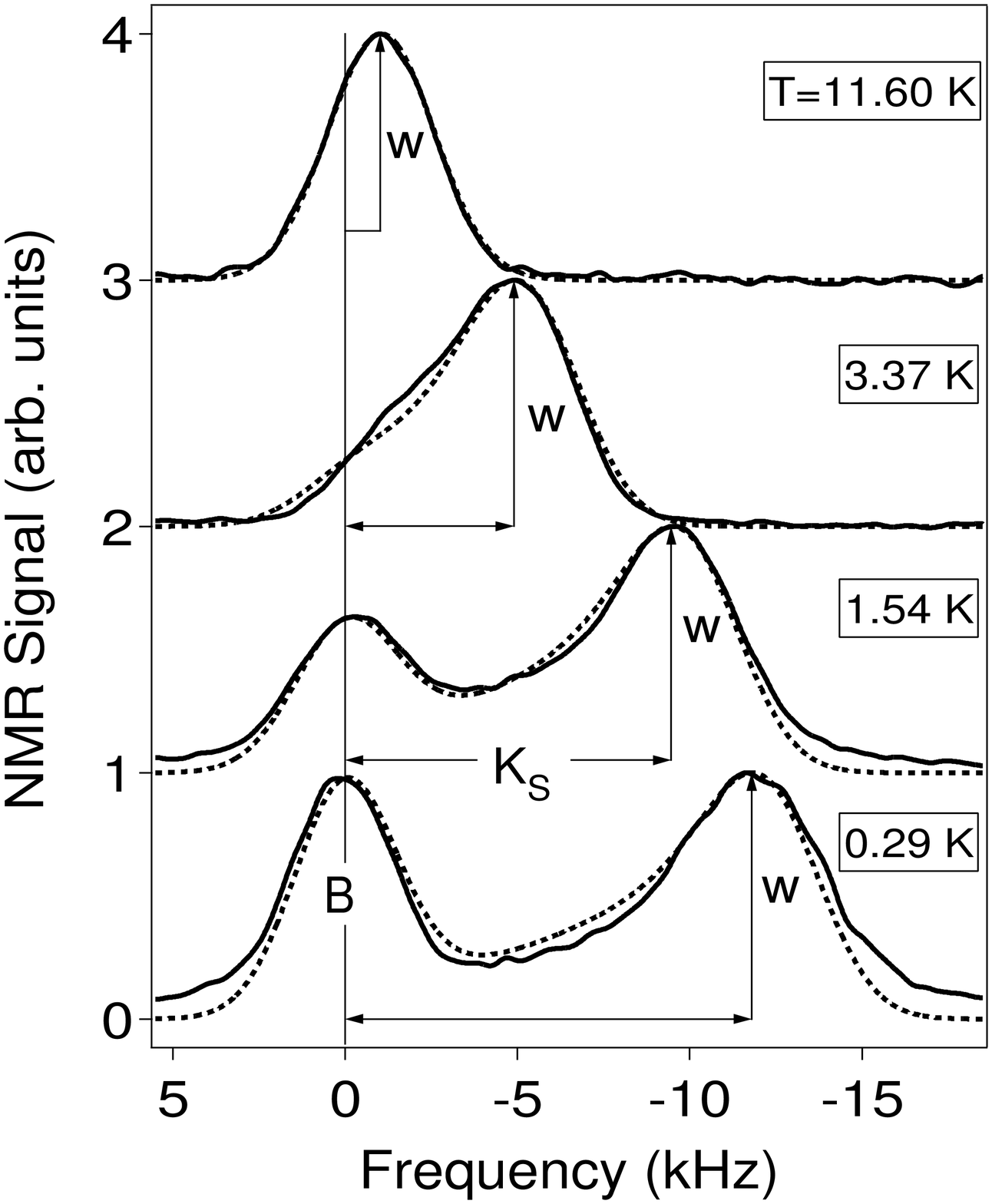}}
\caption{Solid lines: $^{71}$Ga OPNMR spectra 
of sample 10W at 
$\nu$=$\frac{1}{3}$,
taken at $\theta\,$=$\,36.8^{\circ}$
  in $B_{\text{tot}}\,$=$\,12\,$T ($f_{\text{o}}$=155.93 MHz).  
The dashed lines are fits,  described in 
the text. }
\label{fig1}
\end{figure}

 Fig.\,\ref{fig1} shows OPNMR spectra (solid lines) over a range 
of temperatures at $\nu$=$\frac{1}{3}$. Nuclei within
the quantum wells are coupled to the spins of the 2DES via the isotropic 
Fermi contact interaction\cite{slichter}.  The corresponding well
resonance (labeled ``W") is
shifted and broadened relative to the signal from the 
barriers (``B")\cite{opnmr,science}.
We define $K_S$ to be the peak-to-peak splitting between ``W" 
and ``B".  The spectra at $\nu$=$\frac{1}{3}$ are well described by
a simple two-parameter fit\cite{nick} (Fig.\,\ref{fig1}, dashed lines):

\begin{displaymath}    
I(f)=I_{\text{B}}+I_{\text{W}}=
a_{\text{B}}\,g(f)+\!\int^{\textstyle ^{K_{S\text{int}}}
}_{\textstyle _{0}}
\!\!\!\!\!\!\!\!\!\!\!\!\!\!df' 
g(\,f\!-\!f'\,)
\sqrt{\frac{\textstyle f'}{\textstyle K_{S\text{int}}-f'}
}_{\textstyle ,}
\end{displaymath}

where $g(f)$ is a  3.5~kHz FWHM  
Gaussian due to the nuclear spin-spin
coupling\cite{slichter}.
The amplitude of the barrier signal,  $a_{\text{B}}$,
which depends on the OPNMR parameters, 
was suppressed for  small $K_S$ spectra.  The other parameter 
of the fit, the intrinsic hyperfine shift 
of nuclei in the center of each quantum well is 
$K_{S\text{int}}\,$=$\,A_c\:{\cal P}\:n/w$, where
$w$ is the width of the well and $A_c$ is the hyperfine constant.
$K_{S\text{int}}$ can be derived from $K_S$ (both in kHz)
 using the  empirical relation  
$K_{S\text{int}}$=$K_S$$+$$1.1$$\times$$(1$$-$$
\exp(-K_S/2.0))$.
A comparison of $K_{S\text{int}}(T$$\rightarrow\,$0)  in three
different samples yields  
$A_c\,$=$\,(4.5\pm 0.2)\,$$\times\,$$10^{-13}\,$cm$^{3}$/s,
which makes $K_{S\text{int}}$ an {\em absolute} 
measure of the electron spin polarization.  
An implicit assumption in this model is that the
well lineshape is ``motionally narrowed"\cite{slichter}. 
This requires that  
the reversed spins (e.g.\
thermally excited spin waves) are delocalized, so that
$\langle S_{z}(\nu,T)\rangle$, averaged over the NMR time scale 
($\sim\,$40$\,\mu$sec), appears spatially 
homogeneous.

\begin{figure}
\centerline{\epsfxsize=3.55in\epsfbox{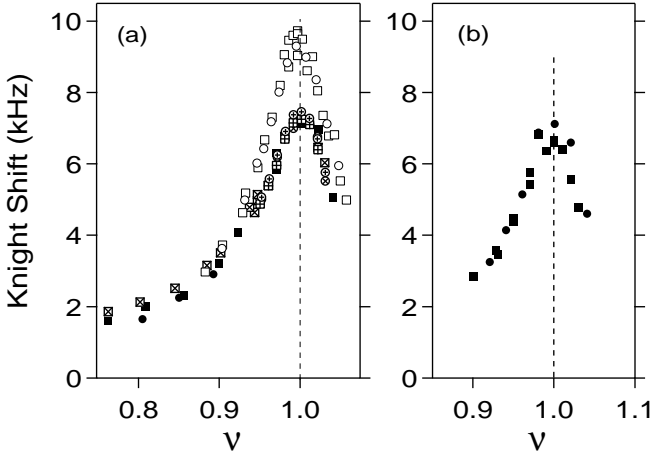}}
\caption{ $K_S(\nu)$
near $\nu$=1 at $T$=1.5 K. (a) Samples 40W (filled symbols,
three separate runs) and
10W (open symbols) at $B_{\text{tot}}$=3.6263 T.
 (b) 40W at $B_{\text{tot}}$=3.2589 T. The densities are
 $n_{\text{40W}}\,$=$\,6.69\,$$\times\,$$10^{10}\,\text{cm}^{-2}$ and
 $n_{\text{10W}}\,$=$\,7.75\,$$\times\,$$10^{10}\,\text{cm}^{-2}$.}
\label{fig2}
\end{figure}

 Using the rotator assembly, we could vary
the angle $\theta$ $(-60^\circ$$<$$\theta$$<$$60^\circ)$
between the sample's growth axis 
 and the applied
 field $B_{\text{tot}}$, thus changing
the filling factor $\nu$=$nhc/eB_{\bot}$ {\em in situ} (here
$B_{\bot}$$\equiv$$B_{\text{tot}}\!\cos\theta$).
  Fig.\,\ref{fig2} shows $K_S$ measurements in the two 
samples near $\nu$=1. The excellent agreement between positive
$\theta$ (squares)
and negative $\theta$ (circles) data is consistent with the rotator
accuracy of $\pm$$0.1^\circ$. We infer the densities $n$ from
these measurements
assuming that $K_S(\theta)$
 peaks at $\nu$=1, hence determining
 $n_{\text{40W}}\,$=$\,6.69\,$$\times\,$$10^{10}\,\text{cm}^{-2}$ and
 $n_{\text{10W}}\,$=$\,7.75\,$$\times\,$$10^{10}\,\text{cm}^{-2}$,
 consistent 
with low-field magnetotransport
characterization of the wafers. These values are very robust, as 
 the four independent runs shown in Fig.\,\ref{fig2} for sample 40W
reproduce $n$ to within $\pm$0.5\%.

Note that the sharp  peak in $K_S$ 
at $\nu$=1 is quite similar to the 
 ``skyrmion feature''  previously observed in a
higher density sample  at stronger $B_{\text{tot}}$
\cite{opnmr}.
The ``size'' of the skyrmion inferred from
Fig.\,\ref{fig2}   ($\tilde{\cal S}$=$\tilde{\cal A}$=3.1 for 
$B_{\text{tot}}\,$$\sim\,$$3.5\,$T)
is slightly
 larger 
than before ($\tilde{\cal S}$=$\tilde{\cal A}$=2.6 for 
$B_{\text{tot}}\,$$\sim\,$$7\,$T)\cite{size},
 in qualitative agreement with the
change in $E_{Z}/E_{C}$\cite{sondhi,fertig}. 
 However, a     
quantitative comparison to the skyrmion model 
will require data below 1.5 Kelvin,
since  $\cal P$($\nu$=1) is only $\sim\,$80$\%$ in Fig.\,\ref{fig2}.

\begin{figure}
\centerline{\epsfxsize=3.55in\epsfbox{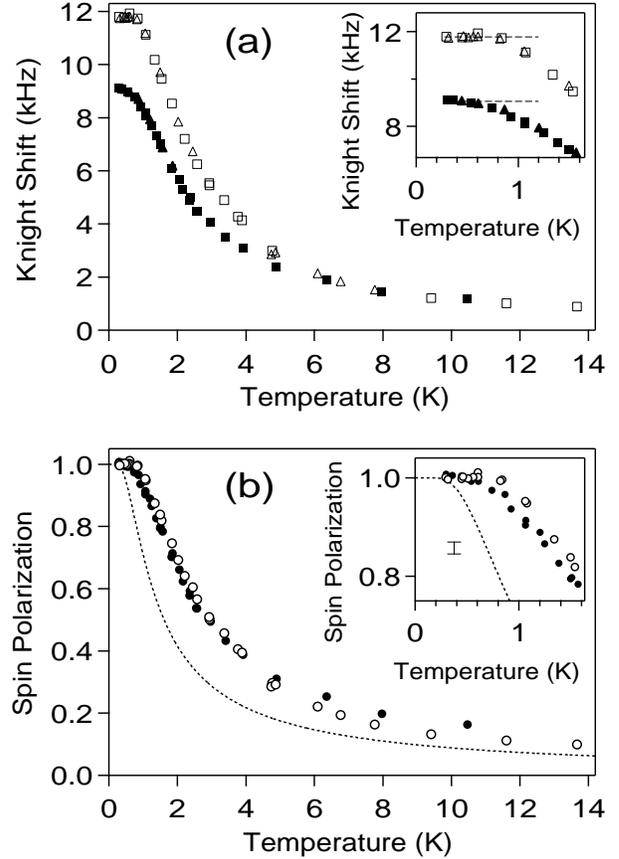}}
\caption{Temperature dependence of (a) $K_S$ 
and (b) $\cal P$ for samples 10W (open
symbols) and 40W  (filled symbols)  at $\nu$=$\frac{1}{3}$
(with $B_{\text{tot}}\,$=$\,12\,$Tesla,
 $\theta_{40\text{W}}$=46.4$^\circ$,
and $\theta_{10\text{W}}$=36.8$^\circ$).
Dashed line is ${\cal P}^{\ast}(T)$, defined in the text. Insets
show the saturation region (note the error bar). 
}
\label{fig3}
\end{figure}

Using the electron densities calculated above, we tilt
each sample by the angle $\theta$ necessary to
achieve $\nu$=$\frac{1}{3}$ in $B_{\text{tot}}\,$=$\,12\,$Tesla
 (where $\theta_{40\text{W}}$=46.4$^\circ$,
 $\theta_{10\text{W}}$=36.8$^\circ$).
 Fig.\,\ref{fig3}(a) shows $K_S$ 
  as a function
 of temperature at $\nu$=$\frac{1}{3}$. Two different symbols
 are used for the 40W data,
 corresponding to independent cool-downs from room temperature,
which demonstrates the reproducibility of the data.   
The inset shows that $K_S$  saturates for both samples at 
low temperatures,
as previously seen at $\nu$=1\cite{opnmr}.
In Fig.\,\ref{fig3}(b) we plot the corresponding
temperature dependence of the electron spin
polarization, using  \mbox{
$ {\cal P}(\nu$=$\frac{1}{3},T)\,$=}$\,
 \frac{K_{S\text{int}}(T)}{K_{S\text{int}}(T\!\rightarrow0)}$.
The resulting curves are almost identical for the two
samples.
The subtle differences that remain
 might be due to a slightly higher spin stiffness\cite{moon}
for sample 10W.

The ${\cal P}(\nu$=$\frac{1}{3},T)$
data in Fig.\,\ref{fig3}(b)  probe 
the neutral spin-flip excitations of a fractional
quantum Hall ferromagnet.  For comparison,
the dashed line is the polarization 
$ {\cal P}^{\ast}(T)$ calculated
for {\em non-interacting} electrons at $\nu$=1, where
$ {\cal P}^{\ast}(T)$=$\tanh(E_{Z}/4k_{B}T)$,
$B_{\text{tot}}$=12\,T, and $g^{\ast}$=\,$-$0.44.  Both 
${\cal P}(\nu$=$1,T)$\cite{opnmr,manfra} and ${\cal P}(\nu$=$\frac{1}{3},T)$
 saturate at higher temperatures than $ {\cal P}^{\ast}(T)$, however,
the data at $\nu$=$\frac{1}{3}$ lie much closer to this 
$ {\cal P}^{\ast}(T)$ limit.
 Fitting   
$\tanh(\Delta/4k_{B}T)$ to  the saturation region of the data,
 we find $\Delta\,$$\approx\,$$2E_{Z}$ at $\nu$=$\frac{1}{3}$, 
as opposed to 
 $\Delta\,$$\approx\,$$10E_{Z}$  at  $\nu$=1\cite{opnmr}.
We also note that the  40W data set is very well described
 by $\Delta\,$$=\,$$1.82\,E_{Z}$ over the {\em entire} temperature range,
 in sharp contrast to the behavior at $\nu$=1.  These results are consistent
with the spin stiffness being much less at $\nu$=$\frac{1}{3}$
 than at $\nu$=1\cite{moon}.
While a recent numerical result\cite{chakra} is in qualitative agreement
with the data in Fig.\,\ref{fig3}(b),
it remains to be seen whether other theoretical 
 approaches, such as
those used at 
 $\nu$=1\cite{skh}, can be modified to explain these data.

\begin{figure}
\centerline{\epsfxsize=3.55in\epsfbox{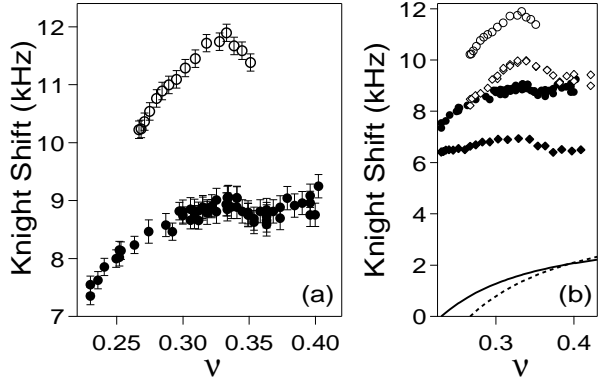}}
\caption{Dependence of $K_S$ on filling factor
at  fixed temperature. Open circles: sample 10W at $T$=0.77\,K,
filled circles: sample 40W at $T$=0.46\,K; open and filled
diamonds: 10W and 40W at $T$=1.5\,K, respectively.
Solid and dashed lines are described in the text.
 }
\label{fig4}
\end{figure}

The Knight shift was also measured at fixed temperature as a function of
sample tilt angle, with $B_{\text{tot}}$=12~T. Fig.\,\ref{fig4}(a) 
shows $K_S(\nu)$ 
near $\nu$=$\frac{1}{3}$ for sample 10W at $T$=0.77\,K,
 and for sample 40W at $T$=0.46\,K.
By these low temperatures, $K_S(\nu$=$\frac{1}{3})$ has 
essentially saturated
for both samples. The data in Fig.\,\ref{fig4}(a) 
show that $K_S(\nu)$ drops on
either side of $\nu$=$\frac{1}{3}$, a result that is reminiscent of earlier
measurements near $\nu$=1\cite{opnmr}.  
The $K_S(\nu$$\sim$$\frac{1}{3})$ feature is distinctly
``sharper'' for sample 10W as opposed to sample 40W. This difference
between the samples is not an artifact of the temperatures plotted, as
Fig.\,\ref{fig4}(b) shows that the distinction is already present by $T$=1.5\,K. 
In order to measure $K_S(\nu)$ this accurately, we took into account the
{\em extrinsic} tilt-angle dependence of the barrier frequency
(Fig.\,\ref{fig4}(b), solid and dashed curves) caused by 
a paramagnetic rotation stage.

\begin{figure}
\centerline{\epsfxsize=3.55in\epsfbox{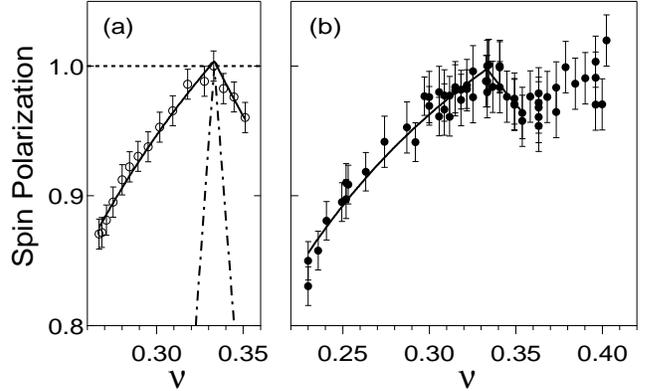}}
\caption{Dependence of $\cal P$  on  filling factor 
at  fixed temperature. (a)  10W  at $T$=0.77\,K (open circles); 
Eq.\,(1)  with $\nu_{o}$=$\frac{1}{3}$ for: 
$\tilde{\cal A}$=$\tilde{\cal S}$=0 (dashed line),
 $\tilde{\cal A}$=0.085 and $\tilde{\cal S}$=0.15 (solid line), and 
 $\tilde{\cal A}$=$\tilde{\cal S}$=1 (dash-dotted line).
 (b)  40W at $T$=0.46\,K (filled circles); Eq.\,(1)  
with $\nu_{o}$=$\frac{1}{3}$, $\tilde{\cal A}$=0.053 and 
$\tilde{\cal S}$=0.10 (solid line). 
}
\label{fig5}
\end{figure}

The $K_S(\nu)$ data shown in Fig.\,\ref{fig4}(a) are converted to the 
corresponding 
electron spin polarization ${\cal P}(\nu)$
$\equiv \frac {K_{S\text{int}}(\nu)}{K_{S\text{int}}(\nu = 1\!/3)}$, and are
plotted in Fig.\,\ref{fig5}.
The polarization of both samples decreases as $\nu$ is varied away from
$\frac {1}{3}$, 
{\em despite the presence of the 12\,T  field!}  Perhaps even more 
remarkably, ${\cal P}(\nu )$ decreases monotonically as $\nu $ is lowered
below $\frac {1}{3}$ over the observed range 
($\frac{\delta \nu} { 1/3} \sim - 30 \% $).
This strongly suggests that the charged quasiparticles and quasiholes of the
$\nu$=$\frac{1}{3}$ ground state involve electron spin flips.

\begin{table}
\begin{tabular}{lrrrrr}
 $\nu\:$$\downarrow$ \space\space  $\backslash$  \space\space  {\em T}$\:\rightarrow$  & 1.5 K & 0.9 K & 0.7 K & 0.5 K & 0.3 K \\ \hline
1/3    &  0 $\%$  & 4 $\%$    & 4 $\%$    &  3 $\%$     & 5 $\%$   \\ \hline
0.29   &  2 $\%$  & 12 $\%$    & 20 $\%$   &  36 $\%$    & 32 $\%$   \\ \hline
0.27   &  12 $\%$  & 21 $\%$   & 45 $\%$   &  69 $\%$    & 53 $\%$   \\ 
\end{tabular}
\caption{The percentage increase of the 
 well linewidth for sample 10W, relative to 
the value of 5.2 kHz at $T$=1.5~K and $\nu$=$\frac{1}{3}$.}
\label{tab1}
\end{table}

A second, independent measurement
 provides further evidence
for the presence of reversed spins
 below $\nu$=$\frac{1}{3}$.
While the high temperature spectra
 are ``motionally narrowed''\cite{slichter},
Table \ref{tab1} shows that the well
 lineshape broadens 
dramatically at low temperatures
 below $\nu$=$\frac{1}{3}$.
This change in the lineshape
 indicates that the time-averaged
$\langle S_{z}\rangle $ is no
 longer spatially homogeneous. 
 The inhomogeneity
requires the existence of spin-reversed
 regions, that become localized 
over the NMR time scale as the temperature
 is lowered below $\sim$0.5\,K
 ($\sim$0.3\,K) for sample 10W (40W)\cite{nick}.
In order to avoid the complication
 of a spatially inhomogeneous 
$\langle S_{z}\rangle $, the data
 presented in Fig.\,5 were 
taken at temperatures that were
 just low enough to saturate
 $K_S(T)$ at $\nu$=$\frac{1}{3}$.

To quantify the rate of
depolarization in Fig.\,\ref{fig5}, we extend a simple model
previously used near $\nu$=1\cite{opnmr}. Our model parametrizes
the effect of interactions in the neighborhood
of a ferromagnetic filling factor $\nu_{o}$$<$1.
 We assume that 
adding a quasiparticle (or quasihole)
to the ground state causes $\tilde{\cal S}$ (or $\tilde{\cal A}$) electron spins
to flip\cite{size}. 
Within this model, the electron spin polarization is:
\begin{equation}    
{\cal P}(\nu) = 1 + 2\Bigl(\frac{1}{\nu}-
 \frac{1}{\nu_{o}}\Bigr)\Bigl( \tilde{\cal S}\,\Theta(\nu\!-\!
 \nu_{o})-\tilde{\cal A}\,\Theta(\nu_{o}\!-\!\nu) \Bigr),
\label{eq1}
\end{equation}
where
$\Theta(x)\equiv\{\,1,\,x\!\geq\!0;\:\text{and}\:\:0,\,x\!<\!0\,\}$.
Using Eq.\,(\ref{eq1}) to fit the data near 
 $\nu_{o}$=$\frac{1}{3}$
(solid lines),
  we find:

$\begin{array}{lll}
	10\text{W:}\:   & \tilde{\cal A}=0.085 \pm 0.005,\:  &   \tilde{\cal S}=0.15 \pm 0.04  \\
	40\text{W:}\:   & \tilde{\cal A}=0.053 \pm 0.008,\:  &   \tilde{\cal S}=0.10 \pm 0.03. \\
\end{array}$

For comparison, the earliest theory\cite{laughlin,prange} of
the $\nu$=$\frac{1}{3}$ ground state assumed
spin-polarized quasiparticles and quasiholes, i.e., 
$\tilde{\cal S}$=$\tilde{\cal A}$=0 (Fig.\,\ref{fig5}, dashed line). Subsequent
calculations \cite{tapash} considered the possibility of
  spin-reversed quasiparticles
 and  quasiholes, i.e.,
$\tilde{\cal S}$=$\tilde{\cal A}$=1 (Fig.\,\ref{fig5}, dash-dotted line).
However, both the early calculations and the more recent 
studies of skyrmion excitations near $\nu$=$\frac{1}{3}$\cite{kamilla,chang}
suggest  $\tilde{\cal S}$=$\tilde{\cal A}$=0 for 
strong magnetic fields.
  On the other hand, our small, non-zero values are
within the bounds set by transport measurements at ambient\cite{haug}
 and high\cite{leadley} pressures.

A much more difficult feature to understand is the fact that our measured 
values are fractional ($\tilde{\cal S}\,$$\sim\,$$\tilde{\cal A}\,$$\sim\,$0.1),
 since the magnetic field should make $\langle S_{z}\rangle$ a good quantum
 number for the N particle system\cite{tapash}.
Of course, our experiment does not
have the resolution to see the effect of adding a single 
quasiparticle to the $\nu$=$\frac{1}{3}$ ground state, thus 
these values for $\tilde{\cal S}\,$ and $\tilde{\cal A}\,$ are the
 {\em average} numbers of 
flipped spins per quasiparticle and quasihole.
 Nevertheless, Eq.\,(\ref{eq1}),
which assumes that all  quasiholes (or quasiparticles) behave in
 exactly the same way, does a remarkably good  job fitting
 our data over the range (0.23$<$$\nu$$<$0.36).  This model is 
expected to break down outside
the ``dilute"
quasiparticle limit (i.e., when $\nu$ gets ``too far" from 
$\frac {1}{3}$), since $\tilde{\cal S}$ and $\tilde{\cal A}$ 
are independent of $\nu$.  
Surprisingly, the above fit actually passes through $\nu$=$\frac{2}{7}$
without modification.  High field magnetotransport measurements on samples
taken from the same wafer as 10W show much more structure, with
 well-developed minima in $\rho_{xx}$ at $\nu$=$\frac{1}{3}, \frac{2}{5},
\frac{2}{7},
$ and $\frac{1}{5}$ at $T\,$=$\,300$~mK\cite{insensitive,willett}.

The possible explanations of these values
 ($\tilde{\cal S}\,$$\sim\,$$\tilde{\cal A}\,$$\sim\,$0.1)
are constrained by many different aspects of the data. 
For example, the values of $\tilde{\cal S}$ and $\tilde{\cal A}$ 
do not appear to change
up to $T$=1.5~K.  
 Furthermore,
the motional narrowing of the NMR line requires that 
the time-averaged electron spin polarization 
is spatially uniform for all $\nu$.

We thank S.\,M. Girvin, A.\,H. MacDonald, N. Read, and S. Sachdev for helpful
discussions. We also thank K.\,E. Gibble, R.\,L. Willett, and K.\,W. Zilm 
for experimental assistance.  This work was
supported by NSF CAREER Grant $\#$DMR-9501925.


\begin{references}
\bibitem{size}Throughout this Letter, we use the convention that
$\tilde{\cal S}$=$\tilde{\cal A}$=0 in the non-interacting limit, instead
of the original definition \cite{opnmr}, ${\cal S}$=${\cal A}$=1, 
 in the same limit.
\end{references}
  \end{document}